\documentclass[
    twocolumn,
	prd,
    amssymb,
	preprintnumbers,
	secnumarabic,
	nofootinbib,
	superscriptaddress]{revtex4-1}

\pdfoutput=1

\usepackage{graphicx}
\usepackage{enumitem}
\usepackage{latexsym}
\usepackage{amsfonts}
\usepackage{amssymb}
\usepackage{pifont}
\usepackage{color}
\usepackage{xcolor}
\usepackage{amsmath}
\usepackage{slashed}
\usepackage{dcolumn}
\usepackage{verbatim}
\usepackage{float}
\usepackage{multirow}
\usepackage{xspace}
\usepackage[normalem]{ulem}
\usepackage{hyperref}
\usepackage{tabularx} 
\usepackage{diagbox}

\definecolor{aquamarine}{rgb}{0.2,0.7,0.6}
\definecolor{mycerulean}{RGB}{0,166,214} 
\definecolor{hypershade}{rgb}{0.3,0.3,0.8}
\definecolor{subtlered}{rgb}{0.8,0.3,0.3}
\hypersetup{
 pdfauthor={Nirmal Raj, Joseph Bramante},
  colorlinks=true,
  citecolor=green,
  urlcolor=purple,
  linkcolor=hypershade
}

\def\Oc{\mathcal{O}}

\newcommand{\beq}{\begin{equation}}
\newcommand{\eeq}{\end{equation}}
\newcommand{\bea}{\begin{eqnarray}}
\newcommand{\eea}{\end{eqnarray}}
\newcommand{\nn}{\nonumber}

\definecolor{rosy}{RGB}{230,235,252}
\definecolor{myframetitle}{RGB}{90,89,170}
\definecolor{myblocktitle}{RGB}{140,185,249}
\definecolor{mytitle}{RGB}{10,80,26}

\definecolor{darkgreen}{RGB}{27,130,45}
\definecolor{darkblue}{rgb}{0,0,0.3}
\definecolor{darkred}{rgb}{0.7,0,0}

\definecolor{light gray}{RGB}{220,220,220}
\definecolor{dark purple}{RGB}{108,0,217}
\definecolor{pink}{RGB}{190,20,100}
\definecolor{orang}{RGB}{193,63,0}
\definecolor{green}{RGB}{11,98,17}
\definecolor{darkpink}{RGB}{153,0,76}
\definecolor{bluegreen}{RGB}{0,102,102}
\definecolor{greenlagan}{RGB}{0,102,0}
\definecolor{redgreen}{RGB}{102,102,0}
\definecolor{Redgreen}{RGB}{153,76,0}
\definecolor{vividviolet}{rgb}{0.62, 0.0, 1.0}
\definecolor{amaranth}{rgb}{0.9, 0.17, 0.31}
\definecolor{palatinateblue}{rgb}{0.15, 0.23, 0.89}
\definecolor{brightpink}{rgb}{1.0, 0.0, 0.5}
\definecolor{cornflowerblue}{rgb}{0.39, 0.58, 0.93}
\definecolor{deepcarminepink}{rgb}{0.94, 0.19, 0.22}
\definecolor{radicalred}{rgb}{1.0, 0.21, 0.37}

\def\TWD{T_{\rm WD}}

\allowdisplaybreaks

\begin{document}

\title{Cosmology of self-replicating universes in black holes \\ formed by dark matter-seeded stellar collapse}
% or natural selection: the dark ending

\author{Joseph Bramante}
\email{joseph.bramante@queensu.ca}
\affiliation{Department of Physics, Engineering Physics, and Astronomy, Queen's University, Kingston, Ontario, K7N 3N6, Canada}
\affiliation{The Arthur B. McDonald Canadian Astroparticle Physics Research Institute, Kingston, Ontario, K7L 3N6, Canada}
\affiliation{Perimeter Institute for Theoretical Physics, Waterloo, Ontario, N2L 2Y5, Canada}

\author{Nirmal Raj}
\email{nraj@iisc.ac.in}
\affiliation{Centre for High Energy Physics, Indian Institute of Science, C. V. Raman Avenue, Bengaluru 560012, India}

\date{\today}

\begin{abstract}
We show that dark matter with certain minimal properties can convert the majority of baryons in galaxies to black holes over hundred trillion year timescales. 
We argue that this has implications for cosmologies which propose that new universes are created in black hole interiors.
We focus on the paradigm of cosmological natural selection, which connects black hole production to a universe's likelihood for existing.
Further, we propose that the universe's timescale for entropy production could be dynamically linked to black hole production in a naturally selected universe. 
Our universe would fit this scenario for models of particle dark matter that convert helium white dwarfs to black holes in around a hundred trillion years, where the dominant source of entropy in our universe are the helium white dwarfs' stellar progenitors, which cease forming and burning also in around a hundred trillion years. 
Much of this dark matter could be discovered at ongoing experiments.
\end{abstract}

\maketitle

%%%%%
\section{Introduction}
%%%%%

Advances in cosmology over the past century have revealed the past of our universe. 
The current epoch of vacuum energy was preceded by matter domination, which began around the time of recombination. 
This in turn was preceded by radiation domination, extending back in time to big bang nucleosynthesis (BBN)~\cite{Planck:2018vyg}. One prevalent model for the era before BBN is cosmological inflation; another possibility is a cosmological bounce, see for example Refs.~\cite{Baumann:2009ds,Brandenberger:2009jq}.

Some authors have proposed that the future evolution of our universe can be used to help us understand its origins. 
The basic idea is to determine whether the dynamics of our universe might lead to the creation of similar universes in the future. 
This line of inquiry can be traced back to seminal thermodynamic investigations: after sufficiently long times a random phase space of trajectories of a closed thermal bath of particles will include fluctuations to lower entropy states~\cite{boltzmann1995lectures,Carroll:2017gkl}, implying that a process of ``recurrence"~\cite{Poincare,QRecurrence,2ndQRecurrence} could keep returning a closed system back to its initial state via random fluctuations.
In light of recurrence, it was conjectured early on~\cite{boltzmann1995lectures,Poincare} that the universe may be a low entropy fluctuation in a long-lived thermal bath of particles. The same idea has arisen in modern cosmology.
 
Cosmological recurrence can occur when a universe fluctuates into existence from a vacuum energy-dominated state~\cite{Garriga:1997ef,Linde:1991sk,Farhi,Farhi:1989yr,Linde:2006nw}, far in the future of our universe. 
Some investigations~\cite{Dyson:2002pf,Albrecht:2004ke,Boddy:2014eba} have focused on obtaining a universe like ours as a likely fluctuation in the far future, using this to question the validity of certain interpretations of quantum mechanics and cosmological paradigms such as chaotic inflation. 
The central tenet of these studies is that our universe should result in more universes like ours.
Put another way, if we consider cosmological evolution forward and backward in time, typical universes produced are those like ours.
This is not strictly required by any manifest physical principle, but for arguments toward that end and a review of cosmological recurrence, see Ref.~\cite{Carroll:2017gkl}. 

One challenge with cosmological recurrence is that there is generally no tangible access to the creation of universes.
It would be really useful if we could take measurements of a rapidly inflating region or observe a patch of spacetime crunch through a cosmological bounce. 
But even if we only have access to inflationary or bouncing patches that are hidden behind a cosmological horizon, already some information may be gained regarding the number of universes being created and the associated dynamics.
Moreover, if we can count the number of universes created behind horizons, this would allow us to begin sorting out the difficult aspects of cosmology and recurrence outlined above.
This brings us to cosmologies where the interior of black holes (BHs) create new universes. 
This could occur via, e.g., a transition to an inflationary phase or a bounce~\cite{Blau:1986cw,Frolov:1989pf,Smolin:1990us,Mukhanov:1991zn}.
Under the assumption that black hole interiors are birth sites of new universes, examining the production of black holes in our universe may teach us about cosmology.

Many of these considerations are contained in the theory of cosmological natural selection (CNS). 
Smolin proposed~\cite{Smolin:1990us} that the dynamics of our universe may be the consequence of small variations in fundamental physical parameters, where these variations occur during the creation of new universes in the interiors of black holes. 
In this theory, our universe's propensity for producing black holes is an expected feature of a {\em typical} universe, since typical universes would have parameters varied (a.k.a. tuned) to produce new universes. 
CNS leads to specific observational predictions and has prompted interesting critiques, which we will discuss in this paper.

\begin{figure}
    \centering
    \includegraphics[width=0.49\textwidth]{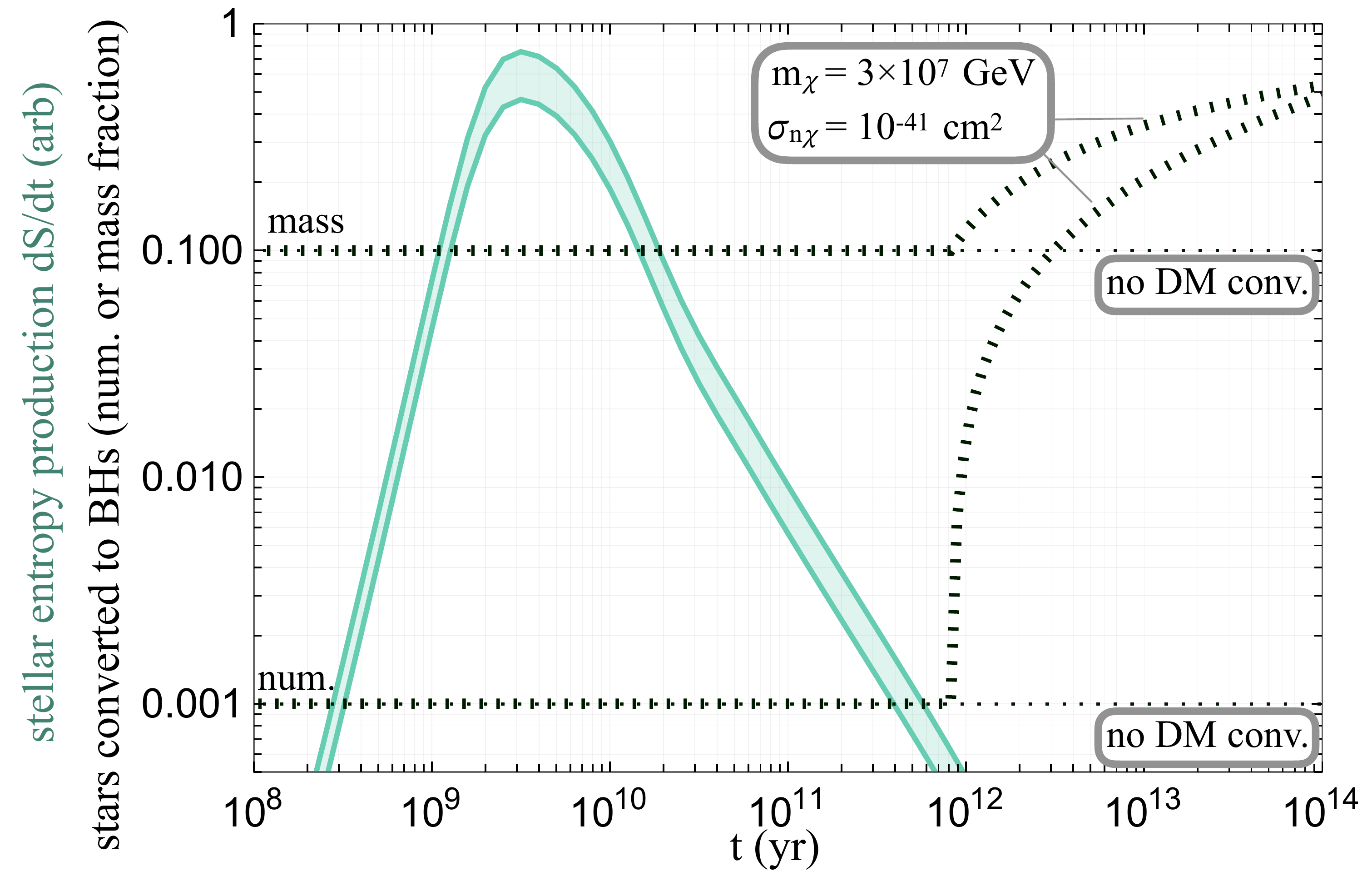} 
    \caption{The total entropy production in our universe with associated uncertainties (solid blue lines) as a function of time, using the historic star formation rate and uncertainties reported in~Ref.~\cite{beacomhopkins}, and extrapolated to future epochs using future star formation~\cite{AdamsLaughlin:1996xe} as detailed in Sec.~\ref{subsec:BHformationrate}. The dashed black curve shows the amount of galactic stellar material that would be turned into black holes for dark matter with the mass and per-nucleon cross section indicated. This result is shown both in terms of the number fraction (labelled ``num.") and mass fraction (labelled ``mass") of stars converted. In the absence of dark matter conversion (``no DM conv."), baryons predominantly form black holes through core collapse supernovae in a roughly constant proportion~\cite{BHmassfunc:Sicilia:2021gtu}. The calculations in Secs.~\ref{subsec:DM-WD}-\ref{subsec:BHformationrate} account for dark matter conversion to black holes of $0.1-0.5~M_{\odot}$ He white dwarfs, which are the predominant site of baryons in galaxies after star formation ceases. Dark matter-induced collapse will also apply to higher mass white dwarfs, neutron stars~\cite{Gould:1989gw, Kouvaris:2007ay, Bertone:2007ae, deLavallaz:2010wp, Kouvaris:2010vv, McDermott:2011jp, Kouvaris:2011fi, Kouvaris:2011gb, selfIDM:Guver:2012ba, Bertoni:2013bsa, Bramante:2013hn, Bell:2013xk, Bramante:2014zca, Bramante:2015cua,Bramante:2016mzo,Bramante:2017ulk,Garani:2018kkd,Kouvaris:2018wnh, Kopp:2018jom, Acevedo:2019gre,East:2019dxt,Tsai:2020hpi,Dasgupta:2020mqg,NSvIR:GaraniGuptaRaj:Thermalizn, Garani:2021gvc,Giffin:2021kgb,Ray:2023auh}, planets~\cite{Acevedo:2020gro}, and other celestial objects, a sub-dominant contribution not included here.}
    \label{fig:starbhform}
\end{figure}

In this work we seek to extend the CNS framework and include the role of dark matter. 
In certain minimal models, particle dark matter would convert the majority of stelliferous baryons into black holes in the coming $10^{14}$ years.
This turns out to be an interesting timescale, as it is when conventional stellar burning comes to an end~\cite{AdamsLaughlin:1996xe}, and is thus also when entropy production in our universe ceases~\cite{BoussoHarnikKribs:2007kq}.
The dynamics of our CNS extension are illustrated in Figure~\ref{fig:starbhform}, which we explain in detail in the rest of this paper.
Altogether, we propose that CNS be extended by setting the end of entropy production in our universe as a relevant dynamical timescale, and by showing that for certain dark matter models this timescale of stellar burning is linked to the summary conversion of most galactic baryons into black holes.

In a nutshell, we propose that
%%%%
\bea
\nn \\
\nn {\rm cosmological \ natural \ selection \ \oplus \ entropy \ dynamics  }  \nonumber \\
\nonumber \\
\nn {\rm suggest \ a \ link \ between \ timescales \ for }  \nonumber \\
\nn {\rm ~1. \ stellar \ entropy \ production, and}  \\ 
\nn \ {\rm  2.\ \ most \ stars \ converting \ to \ BHs.} \\
\label{eq:unific}
\eea
%%%%

In the context of this framework, we obtain predictions for properties of dark matter (DM): if we require that it convert most old white dwarfs (WDs) into black holes over 100 trillion years, this is fulfilled, $e.g.$, by dark matter that has mass between $10^7$ and $10^{11}$ GeV 
and per-nucleon scattering cross section somewhat below weak scale strength, $i.e.$ cross sections around $10^{-42} ~{\rm cm^2}$.
This can be tested in underground experiments with sufficiently large exposures~\cite{APPEC-ARGO:Billard:2021uyg,*DARWINAalbers:2016jon,*XLZDBaudis:2024jnk}, or by astrophysical means like observing the heating of neutron stars (NSs)~\cite{Baryakhtar:2017dbj}.
Frameworks that somewhat depart from the one above can of course be formulated, which often lead to a different set of predictions. 
We discuss some of these in Sec.~\ref{sec:discs}.

Our paper is laid out as follows.
In Section~\ref{sec:execsumm} we provide an executive summary of the cosmological ideas explored in this work.
In Section~\ref{sec:recipe} we more fully review the cosmological paradigms that make up our framework: CNS, entropy dynamics formerly used to construct a so-called causal entropic principle, and the future evolution of stars. 
In Section~\ref{sec:conversion} we incorporate dark matter into this framework, reviewing stellar capture of dark matter and conversion to black holes, and using it to estimate the cosmic formation rate of black holes.
In Section~\ref{sec:discs} we discuss the implications of this work, and particularly re-evaluate criteria for judging CNS.

\section{Summary}
\label{sec:execsumm}
Before proceeding to the main text, we summarize prior cosmological proposals as well as how we use elements of them in our work.
%%%%%%
\begin{itemize}

\item [A.] {\em Cosmological natural selection.}
Reference~\cite{Smolin:1990us} proposed that CNS may occur in our universe through creation of black holes that contain nascent universes, where said black holes are produced predominantly via core collapse supernovae. 
The fundamental parameters of our universe are set through successive, small variation of parameters during the creation of universes inside black holes.
Then the most likely universes we should find ourselves in would be tuned to produce black holes. 

\item [A*.] Here we will propose that most of our universe's CNS-associated black hole production may actually occur through dark matter converting old helium white dwarfs (WDs) into about $0.1~{M_\odot}$ black holes. This can transmute nearly all baryons in galaxies into black holes over 100 trillion year timescales.

\item [B.] Refs.~\cite{Smolin:1994vb,Smolin:1997pe,Smolin:2006gt} argued that 
if one finds that varying a fundamental parameter by a small amount results in more black hole production, CNS is falsified.

\item [B*.] Here we argue that small parametric variations are not a good test of CNS, because a slight apparent increase to the reproductive fitness of the universe does not imply globally improved fitness, which is a conclusion supported by theoretical developments in and experimental tests of evolutionary biology~\cite{Stearns2000}. 
We suggest testing CNS by checking whether the universe's dynamics, which we quantify as being mostly stellar entropy production, are tied to black hole production. 
For certain models of dark matter, black hole production will be dynamically linked to stellar entropy production. 

    \item [C.] {\em Entropy production as a measure of dynamics.} The authors of Ref.~\cite{BoussoHarnikKribs:2007kq} proposed that the cosmological coincidence of currently observed matter-vacuum energy equality can be understood as follows. The rate of observational activity in our universe, which they quantify as being predominantly the rate of stellar entropy production, happens to be maximal over timescales that also maximize the causal volume of our universe. This occurs around the time of matter-vacuum energy equality.  
    
       \item [C*.] Here we do not address the cosmological coincidence problem. 
       However, like Ref.~\cite{BoussoHarnikKribs:2007kq} we will use stellar entropy production to quantify the dynamical action of our universe. In the context of black hole production, we consider the timescale over which most stellar entropy is produced, which is the timescale for $0.1~M_{\odot}$ stars to burn out over 10$^{14}$ years. For certain dark matter models, this is also the timescale over which these stars are converted to black holes.
       
       \item[D.] {\em Future galactic dynamics.} The expected future evolution of stars and galaxies has been detailed in Ref.~\cite{AdamsLaughlin:1996xe}.
       
       \item[D*.] We use the projections in Ref.~\cite{AdamsLaughlin:1996xe} to determine the future star formation rate and the timescale over which $0.1~{M_\odot}$ stars burn out to form helium WDs.
\end{itemize}
In all, we propose to evaluate CNS using stellar entropy production, and show that for certain dark matter models, most stelliferous baryons can be converted to black holes by dark matter over the same timescale that entropy is expended in the universe. Put a different  way, one might say that the burning out of low mass main sequence stars allows heavy dark matter to convert these stars into black holes.

%%%%%%%
%%%%%%%

%%%%%%%%%
\section{Cosmological natural selection and stellar entropy}
\label{sec:recipe}
%%%%%%%%%

We now more fully examine our cosmological proposal's key ingredients.
First we will discuss CNS, next cosmological dynamics quantified by entropy production, then the fate of stellar births and deaths, before presenting our proposal.

%%%%%
\subsection{Cosmological natural selection}
\label{subsec:CNS}
%%%%%

Cosmological natural selection is based on the notion that typical universes are those that reproduce best.
As detailed in Ref.~\cite{Smolin:1990us}, the reproduction of universes is hypothesized to occur via black hole formation, with the assumption that a causally disconnected universe is born beyond the horizon from a bouncing singularity.
(This idea is different from ``the universe as a black hole"~\cite{Pathria1972UnivBH} where an equivalence is suggested between the FRW metric of an internal observer and the Schwarzschwild metric of an external one.)
In each new universe, physical constants slightly vary, presumably due to a dense discretuum of vacua whose minima vary by a small amount due to the bounce dynamics (such as might be seen in something like a string theory landscape).
This results in a sum of universes whose greater part is optimized to create black holes. Ref.~\cite{Smolin:1990us} points out, for instance, that if the neutron-proton mass difference were negative, radiative cooling of neutron (as opposed to hydrogen) gas clouds would be too slow and suppress star formation; if it were higher than the observed 1.29 MeV, nucleosynthesis of elements heavier than hydrogen would be slower, which in turn would subdue the formation of later-generation stars from recycled supernova-ejected stellar material. If it were too high, stable deuterium, and thence heavier nuclides, may never be formed in Big Bang nucleosynthesis. Analogous to the CNS mechanism is Darwinian natural selection, with the slight changes in genetic code paralleling the slight changes in physical constants.

Cosmological natural selection may be contrasted against anthropic selection: the apparent fine-tuning of the cosmological constant (CC) was explained by Weinberg~\cite{WeinbergAnthropicCC:1987dv} by demanding that galaxies (a minimum requirement for observers) exist, and the apparent fine-tuning of the Higgs squared mass was explained by Ref.~\cite{DonoghueAnthropicVEV:1997gf} by demanding that atoms (a minimum requirement for observers) exist.

In Smolin's CNS, the apparent values of fundamental constants are explained by demanding conditions that maximize black hole formation via heavy progenitor stars that end in core collapse supernovae. The conditions required for this to occur happen to overlap with conditions needed for carbon-based life to appear.
An  alternative pathway to realizing the broad paradigm of self-reproducing universes is chaotic eternal inflation~\cite{LindeMultiverseReview:2015edk}.
This has been used as an objection~\cite{SusskindCNS:2004vj} to the underlying principle of CNS, on the basis that if the CC were increased then cosmic self-reproduction via eternal inflation becomes more efficient, whereas we live in a universe with a very small CC.
The essence of CNS, however, is to find a self-reproduction mechanism for the universe that {\em also} explains the apparent fine-tuning making observers possible. 

Since CNS claims that black hole production is maximized in a typical universe such as ours, Smolin also claims that it would be falsified if it can be shown that more black holes form by varying the observed physical constants.
This leads to three predictions by CNS~\cite{Smolin:2006gt}: kaon condensates in NSs, single-parameter inflation, and negligible early star formation.
For the sake of narrative continuity here we relegate a more detailed description of these predictions and our commentary on them to Appendix~\ref{app:CNS3predictions}.

%%%%
\subsection{Quantifying dynamics with entropy}
%%%%

Set forth in Ref.~\cite{BoussoHarnikKribs:2007kq}, the causal entropic principle contends that the fundamental constants, varying over a dicretuum such as the string landscape, are most likely to be near values that maximize the total entropy production in a causally connected region. 
Using this logic, Ref.~\cite{BoussoHarnikKribs:2007kq} is able to predict the measured vacuum energy density $\rho_\Lambda (\simeq 1.25 \times 10^{-123}$ in Planck units).

For $N_{\rm vac}$ vacua with vacuum energy density $< \rho_\Lambda$ and prior probability for these vacua $P$, the anthropically weighted probability distribution for $\rho_\Lambda$ is
%%%
\beq
\frac{dP}{d\log \rho_\Lambda} \propto \rho_\Lambda w(\rho_\Lambda)  \frac{dP}{dN_{\rm vac}} \frac{dN_{\rm vac}}{d\rho_\Lambda}~.
\label{eq:probdistribCC}
\eeq
%%%%
Assuming that all vacua are equally likely before selection effects ($dP/dN_{\rm vac}$ = constant) and that vacua are uniformly distributed near tiny values of $\rho_\Lambda$ ($dN_{\rm vac}/d\rho_\Lambda$ = constant), the causal entropic principle assigns the weights:
%%%%%
\beq
w(\rho_\Lambda) = \Delta S~,
\eeq
%%%%%
where $\Delta S$ is the total entropy produced in the corresponding causal diamond, i.e., the largest space-time region over which matter can interact.
Ref.~\cite{BoussoHarnikKribs:2007kq} finds that the dominant contribution to $\Delta S$ comes from interstellar dust heated by starlight, mostly in the optical range, and emitting it back as more than 100 infrared photons per optical photon.
(A larger contributor to the total entropy is from the horizons of black holes and the causally connected region itself [$i.e.$ de Sitter entropy], but this is neglected as there is no obvious correlation between horizon entropy and observers.)
With this assignment, the probability distribution peaks very close to the measured value of $\rho_\Lambda$, which is within $1 \sigma$ of the measured value. 
The probability distribution in Eq.~\eqref{eq:probdistribCC} is smaller for larger values of $\rho_\Lambda$ because the causal diamonds are smaller, implying less net causally connected entropy production.
It is also smaller for smaller values of $\rho_\Lambda$ because, though the causal diamonds are larger, the entropy production (and hence $w$) is not proportionally larger as star formation slows down and shuts off at late times. 
As a result, $dP/d\log\rho_\Lambda$ is suppressed by the small $\rho_\Lambda$ in Eq.~\eqref{eq:probdistribCC}. 
This approach may be compared with Weinberg's proposed measure for $w(\rho_\Lambda)$, ``observers per baryon".
In spite of setting an excellent upper bound on $\rho_\Lambda$ that is close to the measured value, and hence bringing anthropic considerations of fine-tuning problems into the spotlight, the probability distribution in Eq.~\eqref{eq:probdistribCC} actually peaks at a value of $\rho_\Lambda$ three orders of magnitude higher.

In our work, we do not use the causal entropic principle directly.  
Rather, we only adopt an ingredient from this proposal, which is that the dynamic action of our universe can be quantified in terms of stellar entropy production. Also, we will not place special significance on the time period (around matter-dark energy equality) when the causally connected entropy is maximized, and instead focus on the fact that in our universe most non-horizon entropy is sourced by stars -- where most baryons end up.
It is worth noting that the logic behind the causal entropic principle is not at odds with our work.

%%%%
\subsection{Stellar final mass function}
\label{subsec:FMF}
%%%%

The final ingredient we need for our proposal is the fate of stars: sites of the birth of new universes via conversion to black holes by dark matter.
Star formation is expected to cease in $10^{12}-10^{13}$~years, a timescale set primarily by the depletion of gas in galaxies and secondarily by such complicated effects as gas recycling, gas infall into galactic disks, and uncertainty in modelling of the gas depletion time~\cite{AdamsLaughlin:1996xe}. (Our discussion here is limited to the conventional pathway for star formation via collapse of molecular gas clouds. At around $t = 10^{22}$ years there will be hydrogen-burning stars created via mergers of brown dwarfs~\cite{AdamsLaughlin:1996xe}, but these stars are too few for our consideration.)
For concreteness we mark this end of the Stelliferous Era as $t = 10^{13}$~yr.

Now stellar objects will be in the form of brown dwarfs and compact stellar relics: white dwarfs, neutron stars, and black holes.
Their mass distribution, the ``final mass function" (FMF), may be estimated from the stellar initial mass function (IMF) and the mapping between masses of the progenitor and its corresponding compact object.
For a log-normal form of the IMF, it has been found that the mass fraction of \{brown dwarfs, white dwarfs, neutron stars\} is \{0.10, 0.88, 0.02\}, with black holes making up a sub-percent of the total mass.
We find that for a Kroupa IMF~\cite{KroupaIMF:2002ky}, which captures the low-mass region more accurately, these fractions are \{0.09, 0.90,0.01\}.
In any case we see that most of the baryon mass is in the form of WDs.
Further, the WD mass distribution is weighted toward the lower end, as is the progenitor IMF, and there is a one-to-one mapping between the progenitor mass $M_{\star}$ and WD mass $M_{\rm WD}$~\cite{AdamsLaughlin:1996xe}:
%%%%
\beq
M_{\rm WD} = \frac{M_\star}{1+1.4 M_\star} \exp(M_\star/15)~.
\eeq
%%%%
Thus we find, using a Kroupa IMF, that 77\% of the total stellar object mass (corresponding to 62\% of stellar objects in number) resides in WDs of mass 0.08$-$0.5~$M_\odot$.

%%%%%%
\subsection{Linking entropy and black hole production}
%%%%%%

Using elements of CNS and the entropy production in our universe discussed above, some statements can now be made. 
If there is a mechanism to transform low-mass WDs to black holes soon after their progenitor stars have burned out (over ten-trillion-year timescales), one may state that (1) most of the mass-energy of the entropy producing objects in our universe create new universes, and (2) this conversion happens around the time the universe has concluded its quantifiable dynamics, {\em i.e.}, has produced most of its entropy.
This is reminiscent of biological organisms that reproduce roughly over the timescale of their dynamical entropy increase, expiring soon after (on a logarithmic timescale). 

Guided by these observations, we see that a minimal mechanism exists that could link the dynamics of stellar burnout and black hole production: particle dark matter capturing in stellar remnant WDs and converting them to black holes. 
In Fig.~\ref{fig:starbhform} we show as a function of time the entropy production rate per co-moving volume, superimposed on the fractional mass of WD-converted black holes. 
At $t=10^{14}$ yr we see that both points (1) and (2) above are realized.
The dark matter models that achieve this end are limited, giving predictions for the identity of dark matter and an intriguing target for experimental efforts.

We now propose the following update to CNS in Smolin's original form, which for clarity here we will call CNS1.
In CNS1, about 1\% of baryonic mass ends up inside  black holes (through conventional core-collapse of superheavy stars);
in the new version, more than 75\% of the galactic baryonic mass becomes black holes.\footnote{As discussed later, this number is likely higher once black hole conversion of $> 0.5$ solar mass stars and NSs are accounted for.}
In CNS1 the criteria by which the framework is evaluated is tied to known black hole production mechanisms, but does not tie into the dynamic timescale and production of entropy in our universe; in the CNS proposed here, timescales for entropy production and creation of new universes come out to be about the same, and this is quantified using stelliferous entropy.
Finally, in our proposed reframing of CNS, dark matter plays a central and intricate role in converting stars into black holes.

Next we address the fraction of baryons contained inside galaxies versus in the intergalactic medium. A recent census indicates that less than a quarter of baryonic matter ends up in galaxies~\cite{Nicastro:2018eam}. However, we note that baryons which remain outside galaxies are not producing entropy like those forming stars inside galaxies.

The dynamics for dark matter converting most of the entropy-producing entities (which in our universe is baryons in low mass main sequence stars) into black holes can be summarized using a few equations that estimate the accumulation of dark matter on low-mass He WDs, since most galactic baryons are expected to end up inside them at the end of main sequence burning. 
The next section explores this prediction with a treatment of stellar capture of dark matter, taking into account the formation of He WDs in the future of our universe.

%%%%%%%%%
\section{Conversion of white dwarfs to black holes via dark matter}
\label{sec:conversion}
%%%%%%%%%

In this section, we estimate the timescale over which low-mass helium WDs, end products of low-mass main sequence stars, would be converted into black holes by heavy asymmetric dark matter. 
We assume that dark matter is a Dirac fermion, although as we will see shortly our treatment would apply to bosonic dark matter states as well.
For further aspects of dark matter capture in compact stars see the review in Ref.~\cite{BramanteRajCompact:2023djs}. The mass of Dirac fermionic dark matter most relevant to this study will be $m_\chi \gtrsim 10^6$ GeV. The cosmological production of such heavy asymmetric dark matter has been detailed in a number of prior works such as Refs.~\cite{Benakli:1998ut,Chung:2001cb,Bramante:2017obj,Baker:2019ndr,Kramer:2020sbb,Asadi:2021pwo}, and has been associated with, e.g., moduli decay, other sources of entropy-production or early matter domination, first-order phase-transitions, and non-equilibrium production.

%%%%
\subsection{Dark matter-white dwarf dynamics}
\label{subsec:DM-WD}
%%%%

To determine the amount of dark matter captured by the helium WD, we first consider the total mass flux of dark matter with density $\rho_\chi$ and uniform speed $v_\chi$ passing through the star,
%%%%
\bea
    \dot{M}_{\rm WD} &=& \rho_\chi v_\chi \bigg(\pi R^2_{\rm WD} \bigg[1+\frac{2 G M_{\rm WD}}{R_{\rm WD}v_\chi^2}\bigg] \bigg)  \nonumber \\ 
    &\simeq& 5 \times 10^{27}~{\rm GeV/s}~\bigg(\frac{\rho_\chi}{{\rm GeV/cc}} \bigg)  \bigg(\frac{10^{-3}c}{v_\chi}\bigg) \nonumber \\  && \times \bigg(\frac{M_{\rm WD}}{0.1 M_\odot} \bigg) \bigg(\frac{R_{\rm WD}}{2000 \ {\rm km}} \bigg)~.
    \label{eq:massflux}
\eea
%%%%
This would also be the dark matter mass capture rate in the geometric limit, if multiscatter effects were negligible.
Accounting for the optical thickness of the WD and multiscatter effects, the mass capture rate is
%%%
\begin{equation}
    \dot{M}^{\rm cap}_{\rm WD} = m_\chi \sum_{n=1}^\infty C_n,
   \label{eq:masscapturerate}
\end{equation}
%%%%
where this expression sums over rates for dark matter to be captured after $n$ scatters, given by~\cite{NSMultiscat:Bramante:2017xlb} 
%%%
\begin{align}
C_n &= \pi R_{\rm WD}^{2} p_{n}(\tau) \sqrt{\frac{6}{\pi}} \frac{\rho_{\chi}}{m_\chi v_{\rm \chi}} \Bigg[ (2v_{\rm \chi}^{2} + 3v_{\rm esc}^{2}) \nonumber \\
&\qquad - (2v_{\rm halo}^{2} + 3v_{n}^{2}) \exp \left( -\frac{3(v_{n}^{2} - v_{\rm esc}^{2})}{2v_{\rm \chi}^{2}} \right) \Bigg],
\label{eq:capterms}
\end{align}
%%%%
where  $v_{\rm esc}$ is the WD escape speed, 
$v_n \equiv v_{\rm esc}(1-\beta_{+}/2)^{-n/2}$ the dark matter speed after $n$ scatters with a kinematic factor $\beta_{+}=4m_{\chi}m_{\rm He}/(m_{\chi}+m_{\rm He})^{2}$.
The probability for capture during transit is
%%%%
\begin{equation}
    p_n(\tau) = \frac{2}{n!} \int_{0}^{1} d \cos\theta \left(\cos\theta\right)^{n+1} \tau^n \exp\left(-\tau \cos\theta\right),
    \label{eq:capprob}
\end{equation}
%%%%
where $\tau = 3 \sigma_{\rm T\chi} M_{\rm WD}/(2 \pi R_{\rm WD}^2 m_{\rm He})$ is the optical depth of the He WD.
Finally, $\sigma_{\rm T\chi}$ is the helium-DM scattering cross section, which we take as velocity-independent in this study.
\begin{comment}
   $\sigma_{\rm geo} = 2\pi R^2_{\rm WD} m_{\rm He}/(3 M_{\rm WD})$ is the white dwarf geometric cross section, 
and
$m_{\rm multi} = 2 m_{\rm He}[1 + 2G M_{\rm WD}/(R_{\rm WD} v_\chi^2)]$ is the mass scale of dark matter above which multi-scatter suppression of capture takes effect. 
\end{comment}
For the purposes of capture, we assume that dark matter scatters coherently on nucleons in the helium nucleus, so that for its per-nucleon scattering cross section $\sigma_{\rm n\chi}$ we have  $\sigma_{\rm T\chi} = A^4 F_{\rm H}^2 \sigma_{\rm n\chi}$ in our $m_\chi \gg m_{\rm He}$ region (with A=4). 
The Helm form factor $F_{\rm H}^2(v_{\rm esc})$ that captures loss of nuclear coherence~\cite{Helm:1956zz} is set to unity here; we have checked that this is accurate for helium WDs, which have both smaller nuclei and lower escape speeds than, $e.g.$, carbon-oxygen WDs~\cite{Acevedo:2019gre,Acevedo:2023xnu}. However, we will restore the Helm form factor to determine energy deposition when the dark matter collapses to form a small black hole.

Unlike in WDs made of heavier elements, the interiors of helium WDs do not crystallize due to weak Coulomb binding~\cite{HeWDAlthausBenvenuto1997}.
This considerably simplifies the treatment of their passive cooling, as it never enters the non-trivial Debye regime.
The time for a WD to cool to an internal temperature $\TWD$ ($\ll$ its initial temperature), independent of the WD mass, is~\cite{ShapiroTeukolsky} 
%%%%%
\bea
t_{\rm cool, He-WD} &=& 10^{13}~{\rm yr}~\bigg(\frac{2.7\times10^5}{\TWD}\bigg)^{5/2}~.
\label{eq:tcoolpreDebye}
\eea
%%%%
When treating dark matter-white dwarf dynamics, we will take $\TWD = 3\times 10^5$~K as is relevant to the timescale of interest.
The captured dark matter will scatter and rescatter with the interior of the He WD, forming a ``thermalized'' sphere in virial equilibrium at its center, with radius~\cite{Acevedo:2020gro}
%%%%
\bea
 \nn r_{\rm th} &=& \sqrt{\frac{9 \TWD}{4 \pi G \rho_{\rm WD} m_\chi}} \\
\nn &=& 0.2~{\rm km}\left(\frac{10^7~{\rm GeV}}{m_\chi}   \frac{\TWD}{3 \times 10^5~{\rm K}}  \frac{5\times10^4~{\rm g/cm^3}}{\rho_{\rm WD}} \right)^{1/2}~,\\
\eea
%%%%
where we have normalized the WD central density to the one corresponding to 0.1~$M_\odot$ mass with a Salpeter equation of state~\cite{HeWDDensityMathewNandy2017}.

From the energy loss rate of dark matter of energy $E$,
%%%%
\beq
\frac{dE}{dt} = \frac{-2 \rho_{\rm WD} A^4 \sigma_{\rm n\chi} v_{\rm rel}E}{m_\chi}~,
\label{eq:Elossrate}
\eeq
%%%%
with $v_{\rm rel} \simeq \sqrt{2E/m_\chi}$ the dark matter-helium relative velocity, 
the time taken by dark matter to thermalize with the WD core is~
%%%%
\bea
\nn t_{\rm therm} &=& \frac{1}{4\sqrt{3}} \bigg(\frac{m_\chi^3}{\TWD}\bigg)^{1/2}\frac{1}{\rho_{\rm WD}A^4\sigma_{n\chi}}~\\
&=&4.4\times 10^4~{\rm yr} \bigg(\frac{m_\chi}{10^7~{\rm GeV}} \bigg)^{3/2} \bigg(\frac{3\times10^5~{\rm K}}{\TWD} \bigg)^{1/2} \nn \\
&& \bigg( \frac{5\times 10^4~{\rm g/cm}^3}{\rho_{\rm WD}}\bigg) \bigg(\frac{10^{-40}~{\rm cm}^2}{\sigma_{\rm n\chi}}\bigg)~.
\label{eq:ttherm}
 \eea
%%%%

Once enough dark matter accumulates inside this central region, it will become unstable to small perturbations and collapse, so long as it satisfies the Jean's length criterion and is self-gravitating~\cite{Acevedo:2020gro}.
The mimimum mass to accumulate for self-gravitation is
%%%%
\begin{align}
  &  M_{\rm sg}   = \frac{4\pi}{3} \rho_{\rm WD} r^3_{\rm thermal} \nonumber \\ &\simeq 10^{42}~{\rm GeV} \left( \frac{10^7~{\rm GeV}}{m_\chi} \cdot \frac{\TWD}{3\times 10^5~{\rm K}} \right)^{3/2} \left( \frac{5 \times 10^4~{\rm g/cm^3}}{\rho_{\rm WD}} \right)^{1/2}~.
  \label{eq:Msefgrav}
\end{align}
%%%%
To form a black hole of Schwarzschild radius $r_{\rm Schw}$, the time for collapse from an initial energy $GM_{\rm sg}m_\chi/r_{\rm th}$ to a final energy $GM_{\rm sg}m_\chi/r_{\rm Schw}$ is obtained from the energy loss rate in Eq.~\eqref{eq:Elossrate} as
%%%%%
\beq
t_{\rm collapse} = \frac{1}{4\sqrt{2}}\frac{m_\chi}{\rho_{\rm WD}A^4\sigma_{\rm n\chi}}~.
\label{eq:tcollapse}
\eeq
%%%%%

During the process of collapsing to form a small black hole, the dark matter sphere could potentially ignite helium thermonuclear fusion and spark a Type Ia-like supernova. 
We will find that this does not occur for dark matter collapse in He WDs, in contrast to carbon-oxygen WDs where it can occur (and may indeed be occurring for some of the dark matter parameters we consider)~\cite{Bramante:2015cua,Acevedo:2019gre,Raj:2023azx,Acevedo:2023cab}.
This is because the gravitational energy shed by dark matter is diffused away by the He WD material faster than the rate at which it is injected by dark matter scattering.
This implies runaway fusion is not sustained in He WDs, as opposed to in C-O WDs with denser cores and different heat conductivities~\cite{Bramante:2015cua,Acevedo:2019gre}.
The energy transferred to the He WD by the collapsing dark matter sphere is~\cite{Acevedo:2019gre}
\begin{align}
    \dot{Q}_{\chi} = \frac{M_{\rm sg} \Delta E}{m_\chi t_{\rm N\chi}},
\end{align}
where $t_{\rm N\chi} \equiv (n_{\rm He} \sigma_{\rm T \chi} v_{\rm vir})^{-1}$, 
and in the $m_\chi \gg m_{\rm He}$ limit $\Delta E \simeq m_{\rm He} v_{\rm vir}^2$, with $v_{\rm vir}$ the virial speed. 
We evaluate the sphere's energy injection at $v_{\rm vir} \simeq 0.02$, which is where the energy input is maximized before Helm form factor suppression kicks in~\cite{Acevedo:2019gre}. For $0.1 \to 0.5~M_\odot$ He WDs we find corresponding heating rates of
\begin{align}
    \dot{Q}_{\chi} &= \left( 3 \times 10^{25} \to 10^{26}\right) {\rm GeV/s}  \nonumber \\ & ~~~~~~~~ \times \left(\frac{\sigma_{\rm n \chi}}{10^{-41}~{\rm cm^2}} \right) \left(\frac{3\times10^7~{\rm GeV}}{m_\chi} \right)^{5/2}.
    \label{eq:Qdotchi}
\end{align}

During the dark matter sphere's collapse, the heat diffusion rate will be determined by the conductive opacity of the He WD material, obtained using the conductivities of solid helium in Ref.~\cite{PotekhinUrl}. To ignite the core of $0.1 \to 0.5~M_\odot$  He WDs requires a trigger temperature $T_{\rm trig} = 7 \times 10^8~{\rm K}$ and a minimum mass of ignited material of $10^{14} \to 10^{10}$ grams~\cite{TimmesHe}. The diffusion rate for $0.1 \to 0.5~M_\odot$ He WDs is then~\cite{Acevedo:2019gre}
\begin{align}
    \dot{Q}_{\rm diff} = \frac{4 \pi^2 T_{\rm trig}^3(T_{\rm trig}-T_{\rm WD})r_{\rm trig}}{15 \kappa \rho_{\rm WD} } \simeq 10^{32} \to  10^{30}~ {\rm GeV/s},
    \label{eq:Qdotdiff}
\end{align}
where the trigger radius $r_{\rm trig}$ is determined from the trigger masses quoted above, and $\kappa$ is the thermal conductivity.
We note that diffusion out of the ligher WDs is faster, since they have larger trigger mass and trigger radius. By comparing Eqs.~\eqref{eq:Qdotchi} and \eqref{eq:Qdotdiff} we conclude that it is not possible to ignite our He WDs during the collapse of the dark matter sphere.

Past collapse, to form a black hole we require that the accumulation of dark matter exceed its Chandresekhar mass\footnote{For spin-0 dark matter with small self-interactions, the Chandrasekhar condition is relaxed to $M_{\rm Ch} \approx M_{\rm Pl}^2/m_\chi$, however a black hole must form via a Bose-Einstein condensate, whose dynamics before and after collapse requires more detailed investigation~\cite{BramanteRajCompact:2023djs}.},
%%%%
\bea
    M_{\rm Ch} \approx\frac{M_{\rm Pl}^3}{m_\chi^2} 
 \simeq 2 \times 10^{43}~{\rm GeV}  \bigg( \frac{10^7~{\rm GeV}}{m_\chi} \bigg)^2~.
 \label{eq:MCh}
\eea
%%%%

Once a black hole of mass 
\beq
M_{\rm BH} = \max[M_{\rm sg},M_{\rm Ch}]
\label{eq:MBH}
\eeq
is formed from the dark matter agglomerate, the net rate of its growth,
%%%%
\beq
\dot{M}_{\rm BH} = \frac{4\pi (G M_{\rm BH})^2 \rho_{\rm WD}}{c_s^3 }+  \dot{M}_{\rm WD}^{\rm cap}~- \frac{1}{15360\pi (G M_{\rm BH})^2},
\eeq
%%%%
is determined by its Bondi accretion of the WD material (the first term in the right-hand side, with $c_s \simeq 0.003$ the sound speed of the WD material as obtained from the ideal degenerate Fermi gas EoS in Ref.~\cite{HeWDDensityMathewNandy2017} which is numerically close to the Salpeter EoS),
the dark matter mass capture rate (second term), and
the rate of the Hawking evaporation of the black hole (third term).
If the first term dominates, the timescale for Bondi accretion is obtained as 
%%%%
\beq
t_{\rm Bondi} = \frac{c_s^3}{4\pi G^2 M_{\rm BH} \rho_{\rm WD}}~.
\label{eq:tBondi}
\eeq
%%%%
If the third term dominates, the evaporation timescale is obtained as
%%%
\beq
t_{\rm evap} = 5120 \pi G^2 M^3_{\rm BH}~.
\label{eq:tevap}
\eeq
%%%%
We now comment on the timescales for dark matter to accumulate in and convert compact stars into black holes. First, we note that the above expression for the evaporation timescale should break down at sufficiently small black hole mass, since this introduces a high-energy regime of black hole evaporation which is still being studied \cite{Page:2004xp}. Secondly, it will be useful for future numerical simulations of dark matter in compact stars to validate or improve on the above timescale estimates. One such study has been conducted for a small black hole formed from dark matter inside a neutron star, which found reasonable agreement with the late-stage millisecond-long Bondi accumulation time predicted by the above formulae \cite{East:2019dxt}.

%%%%%%%
\begin{figure}
    \centering
    \includegraphics[width=0.47\textwidth]{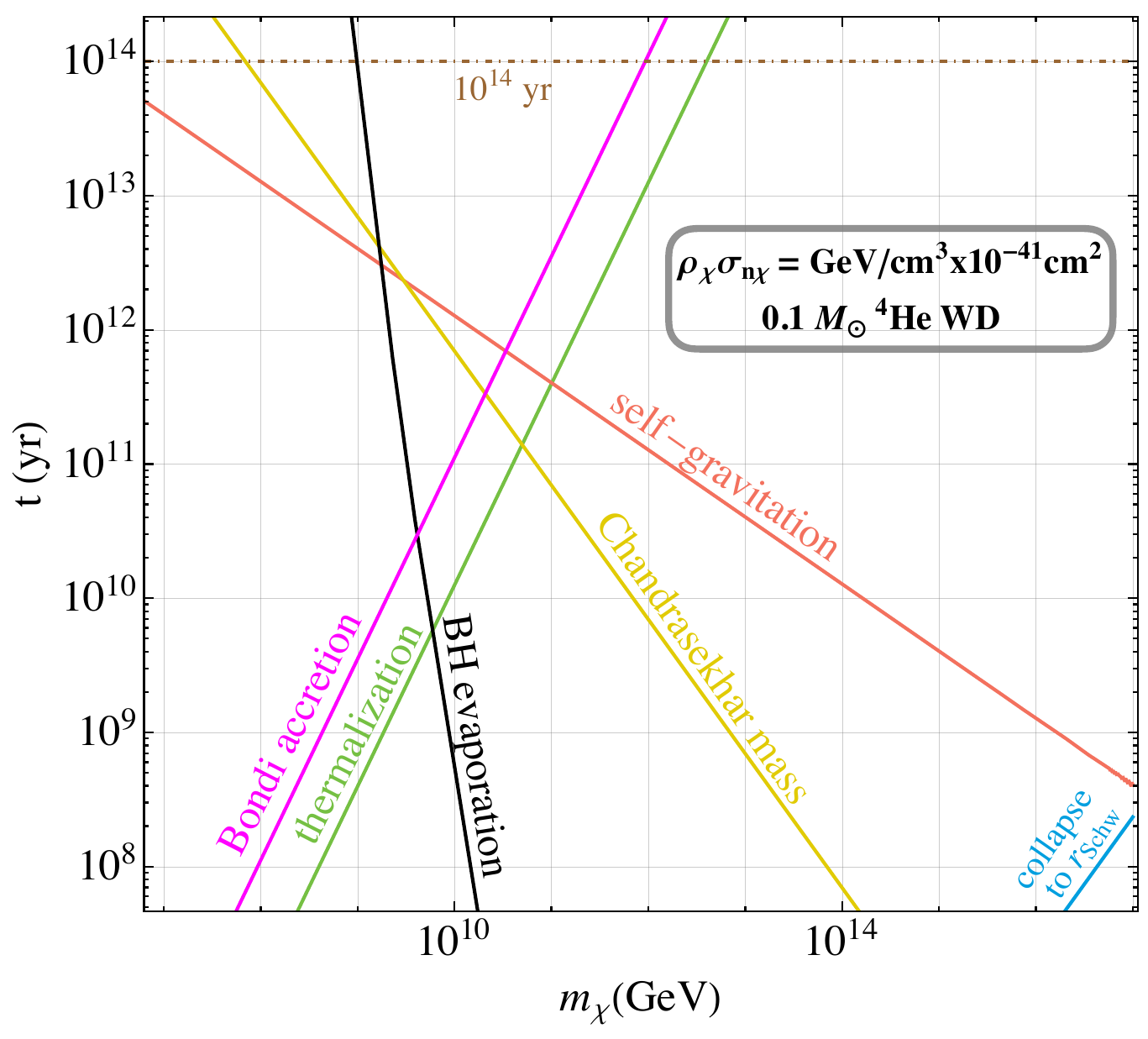} \ 
       \caption{Timescales for various processes to occur as a function of dark matter mass in a 0.1 $M_\odot$ helium white dwarf for an ambient dark matter density of GeV/cm$^3$ and speed of 10$^{-3}~c$.
       For comparison is shown $t = 10^{14}$ yr, a time after which all main sequence stars will have burnt out, which is also the time required for dark matter to convert helium white dwarfs to black holes.
       The benchmark dark matter-nucleon elastic scattering cross sections $10^{-41}$~cm$^2$ is used for illustration. 
       To form a black hole in the interior of the white dwarf, dark matter must first thermalize with the stellar material, then self-gravitate, and collapse to the Schwarzschild radius of the self-gravitating mass.
       This central black hole would simultaneously Bondi-accrete white dwarf material and Hawking-evaporate.
       For cross sections around 10$^{-41}$ cm$^2$ and dark matter masses $\Oc(10^7 - 10^9)$~GeV, the central black hole consumes the parent star within $10^{14}$~yr, consistent with Figure~\ref{fig:starbhform} that integrates over 0.1$-$0.5 $M_\odot$ helium white dwarfs. 
       See Sec.~\ref{subsec:DM-WD} for further details and the text below Eq.~\eqref{eq:tevap} for the equations used to plot these curves.}
    \label{fig:timescales}
\end{figure}
%%%%%%%

The timescales for these processes to occur as a function of the dark matter mass in a $3 \times 10^5$~Kelvin 0.1 $M_\odot$ helium WD are shown in Fig.~\ref{fig:timescales}.
These are for dark matter {thermalization}: Eq.~\eqref{eq:ttherm}, 
{self-gravitation}: $M_{\rm sg}/\dot{M}^{\rm cap}_{\rm WD}$ from Eqs.~\eqref{eq:masscapturerate} and \eqref{eq:Msefgrav},
{collapse} to Schwarzschild radius: Eq.~\eqref{eq:tcollapse},
attainment of {Chandrasekhar mass}: $M_{\rm Ch}/\dot{M}^{\rm cap}_{\rm WD}$ from Eqs.~\eqref{eq:masscapturerate} and \eqref{eq:MCh},
and finally {Bondi accretion} by and {evaporation} of the black hole formed: Eqs.~\eqref{eq:tBondi} and \eqref{eq:tevap}.
These are plotted for a dark matter-nucleon cross section of $10^{-41}$~cm$^2$, assuming an ambient dark matter density of GeV/cm$^3$ and average speed of $10^{-3}~c$. 
These parameters are those of a typical region in a typical galaxy~\cite{SilkMamonReview}. 
For comparison is also shown the $t = 10^{14}$~yr timescale.

We see that within 10$^{14}$ yr dark matter with $m_\chi < \Oc(10^{12})$~GeV can thermalize with He WD cores, collapse in a much shorter timescale, and  if $m_\chi > \Oc(10^7)$~GeV could have accumulated enough to form a black hole.
These values are consistent with Figure~\ref{fig:starbhform}, which integrate over 0.1 to 0.5~$M_\odot$ He WDs. 
For $m_\chi < 10^9$~GeV the dark matter-induced black hole Bondi-accretes the surrounding material and consumes the host WD.
In this scenario all low-mass WDs are converted to black holes, implying most of the baryonic mass of the universe is turned into the mass of the cradles of new universes. 
For $m_\chi >$ a few giga-GeV, the black hole evaporates at a rate faster than the Bondi accretion rate, so the WD is not destroyed.
But this means that black holes are formed sequentially in rapid succession in the host, implying that the number of new universes is increased.
From the discussion in this sub-section the scaling of the various timescales with  $\sigma_{\rm n\chi}$ may be inferred.
In particular, for smaller cross sections black hole formation takes much longer.
For instance, for $\sigma_{\rm n\chi} = 10^{-45}$~cm$^2$ and $m_\chi$ in the vicinity of $10^{10}$~GeV the timescales for dark matter thermalization, self-gravitation or Chandrasekhar mass formation coincide at around $10^{16}$~yr; Bondi accretion of the WD or evaporation follows quickly.
While possible, this is not as interesting as the case of larger cross sections where the black hole formation occurs at $10^{14}$~yr, the timescale for maximal entropy production of the universe.   

Now as WDs outnumber NSs by only a factor of about 100 in the FMF, one may wonder if NSs host more sequentially formed black holes than low-mass WDs over a given interval of time.
This could happen if the Hawking evaporation of black holes in NSs is more than 100 times faster than in our WDs.
However, for the range of dark matter masses where this could potentially occur, the scenario is pre-empted by the fact that the black hole mass in Eq.~\eqref{eq:MBH} is the dark Chandrasekhar mass~$M_{\rm Ch}$, which is the same for both WDs and NSs, so that the evaporation timescale is the same for both.
While in WDs the self-gravitating dark matter mass $M_{\rm sg}$ and $M_{\rm Ch}$ are comparable, in NSs $M_{\rm sg} \ll M_{\rm Ch}$ for $m_\chi < 10^{27}$ GeV due to the relatively enormous densities of $10^{15}$~g/cm$^3$ versus $10^5$~g/cm$^3$ (see Eq.~\eqref{eq:Msefgrav}) and the small $\Oc(10^2-10^3)$~Kelvin NS temperatures~\cite{coolingminimal:Page:2004fy,NsvIR:otherinternalheatings:Reisenegger,NSreheated:RajShivanna:2024kjq}.
So for dark matter particles taken to be elementary and sub-Planckian, the dark matter-induced black hole dynamics are determined by the Chandrasekhar criterion.

We note that the results shown in Figure \ref{fig:timescales} do not include contributions from dark matter converting NSs and other celestial objects to black holes. Including this transmuted component in future work would tend to increase the number and mass of black holes formed by dark matter. In addition, there are effects unaccounted-for in future star formation~\cite{AdamsLaughlin:1996xe} arising from an increase in galactic gas metallicity.
This could be triggered by nucleonic interactions of heavy dark matter that would ignite all C-O WDs over the long timescales we consider~\cite{Bramante:2015cua,Acevedo:2019gre,Janish:2019nkk,Acevedo:2023cab}.
The fractions of $e.g.$ C, O, and Fe in galactic gas would then increase. These C-O WD explosions might also change galaxy properties~\cite{Acevedo:2023cab}. However, since most of the progenitor stars making the future He WDs we are studying have already formed, we believe the results we have obtained here should be robust against such astrophysical refinements.

%%%%
\subsection{Entropy production and black hole formation}
\label{subsec:BHformationrate}
%%%%

To estimate the integrated entropy production rate shown in Figure \ref{fig:starbhform} at time $t$ we follow Ref.~\cite{BoussoHarnikKribs:2007kq}, with some modifications. 
We use the Kroupa IMF~\cite{KroupaIMF:2002ky} given in Sec.~\ref{subsec:FMF}, $\xi_{\rm IMF} (M_\star)$, which weights the entropy production rate per star, $d^2 s/dN_\star dt = L_\star/T_{\rm eff}$, where the main sequence star luminosity $L_\star \propto M_\star^{3.5}$ and the temperature of reprocessed starlight $T_{\rm eff} = 20$~meV.
This is then convolved with the redshift-dependent star formation rate, $\dot \rho_\star$, and integrated over the birth times $t^\prime$ of stars born prior to $t$ and volume $V_c$:
%%%%
\bea
\nn  \frac{dS}{dt} (t) = \int dV_c \int_0^t dt^\prime &&\int_{0.08 ~M_\odot}^{M_{\rm max} (t-t^\prime)} \frac{1}{\langle M \rangle} dM_\star  [ \xi_{\rm IMF} (M_\star) \\
&& \times \dot{\rho}_\star (t^\prime) \frac{d^2s}{dN_\star dt}(M_\star)]~.
\eea
%%%%
Here $\langle M \rangle$ is the IMF-weighted average initial mass, and the SFR $\dot{\rho}_\star = d^2 M_\star/dV_c dt$ is taken as the fit to the measured star formation rate in Ref.~\cite{beacomhopkins} which is then matched to an exponential-in-time decay function as expected for future star formation in gas-depleted galaxies~\cite{AdamsLaughlin:1996xe}. 
We also use a main sequence stellar lifetime $\simeq 10^{10}~{\rm yrs}~ (M_\odot/M_\star)^{2.5}$, which determines the upper limit of the third integral.

The black hole production rate from stellar conversion by dark matter-seeded collapse is given by
%%%%%
\bea
\nn \frac{dN_{\rm BH}}{dt} (t) = \int dV_c \int_0^t dt^\prime && \int_{0.08 M_\odot}^{100 M_\star}  \frac{1}{\langle M \rangle} dM_\star  [ \xi_{\rm IMF} (M_\star) \\
&& \times \dot{\rho}_\star (t^\prime) \frac{d^2 M_{\rm BH}}{d M_\star dt} (t-t^\prime)]~, 
\label{eq:BHformationrate-number}
\eea
%%%%
where again we use the Kroupa IMF and
%%%%
\beq
\frac{d^2 M_{\rm BH}}{d M_\star dt} (t-t^\prime) = 
\begin{cases} f_{\rm rem}(M_{\star}), \  t-t^\prime > t_{\star \to \bullet} \\ 
 0, \ \ \ \ \ \ \ \ \ \ \ t-t^\prime < t_{\star \to \bullet}~,
\end{cases}
\eeq
%%%%
where $f_{\rm rem}$ is the differential of the mass of the black hole formed to the progenitor mass $M_\star$, and $t_{\star \to \bullet}$ is the time taken for black hole formation.
For the dark matter capture-induced black hole formation mechanism treated here, $t_{\star \to \bullet}$ is determined by computing the time required to convert $0.1-0.5$~{$M_\odot$} He WDs to black holes, as detailed in Sec.~\ref{subsec:DM-WD} and $f_{\rm rem} (M_\star) = 1$. 
For the usual astrophysical route to black hole formation, $t_{\star \to \bullet}$ is the lifetime of the star and $f_{\rm rem}$ is as given in Ref.~\cite{BHmasspectrum-Spera-2015}.
The net {\em mass} rate of black hole formation is given by dropping the factor of $\langle M \rangle^{-1}$ from Eq.~\eqref{eq:BHformationrate-number}.
Here we do not include black hole formation from binary mergers of stellar systems as it is a small effect~\cite{BHmassfunc:Sicilia:2021gtu}.

\begin{table*}
\caption{Variations of the cosmological natural selection framework studied in this work. See Sec.~\ref{subsec:variations} for a detailed discussion.}
\centering
\renewcommand{\arraystretch}{1.5} 
\begin{tabularx}{\textwidth}{ p{1.7cm} p{2.5cm} XX} 
\hline\hline
\backslashbox{\textcolor{mycerulean}{agent}}{\textcolor{subtlered}{measure}} & & \textcolor{subtlered}{net black hole mass} & \textcolor{subtlered}{black hole count} \\ 
\hline
\multirow{2}{*}{\textcolor{mycerulean}{matter mass}} 
  & \textcolor{aquamarine}{baryons}               & {\bf this work} (heavy particle DM)           & {\bf this work} (super-heavy particle DM) \\
  & \textcolor{aquamarine}{dark matter}           & PBHs, or DM structures forming BHs~\cite{mirrorNSs:Hippert:2021fch,axinovae:Fox:2023xgx}                & very light PBHs~\cite{PBHslight:Bai:2019zcd,PBHslight:Lehmann:2019zgt} \\
\hline
\textcolor{mycerulean}{dark energy} & & \multicolumn{2}{c}{Vilenkin's dS universe~\cite{CNSdS:Vilenkin:2006hq}} \\ 
\hline\hline 
\end{tabularx}
\label{tab:othermeasures}
\end{table*}

%\newpage
%%%%
\section{Discussion}
\label{sec:discs}
%%%%

We have outlined a theory of cosmological natural selection, where the dynamical timescale for stars to burn out in our universe is linked to the production of new universes, through dark matter-seeded black hole collapse in old helium white dwarfs. As in Ref.~\cite{Smolin:1990us}, we assume that the dynamic which has shaped our universe into this format involves fundamental constants varying to some degree when new universes are created inside black holes. This naturally leads to the question, ``But why doesn’t our universe create more black holes faster, by varying its parameters to some other values?'' In Appendix~\ref{app:CNS3predictions} and the remainder of this discussion, we respectively review and expand on this question, which has been explored previously~\cite{Smolin:1990us,Smolin:1994vb,Smolin:1997pe,CNSdS:Vilenkin:2006hq,Smolin:2006gt}.

Here let us introduce a different perspective on this question, drawn from the life history theory for biological evolution~\cite{Stearns2000}. 
Experimental studies of complex biological organisms have shown that attempting to ``increase the fitness'' of organisms by naively dialing a parameter to increase offspring will often lead to \emph{fewer} grandchildren, in a suitably adapted system.  As a concrete example, the number of eggs in kestrel clutches was both increased and decreased~\cite{kestrel}, and it was found that the resulting grandchildren populations diminished in both cases. While this may seem counter-intuitive at first, it is important to keep in mind that in complex systems governed by a large number of selectively varied parameters, there are non-obvious trade-offs between such factors as resource gathering, system growth, and reproduction. 

Moreover, in the context of varying the parameters governing the behavior of a complex evolving system, the pathways available for parametric variation play a role in determining the present and future fitness of the system. Applying lessons from evolutionary-developmental biology~\cite{arthur2002emerging}, it may be the case that by naively shifting a fundamental constant of our universe that results in an increase in black hole production, we are now in a location on our landscape of parameters where further variations will dramatically {\em decrease} black hole production. Put another way, a near-term variational gain in black hole production may imply a much larger future variational diminution of black holes produced.

Faced with this, here we have proposed as a test of CNS, to simply examine the predominant quantifiable dynamical aspect of our universe, which we take to be stellar entropy production, and see if it has any intrinsic connection to the putative method of new universe production -- here, black hole production. We have seen that for certain kinds of dark matter these two phenomena could be connected. So far we have focused on what appears to be the simplest scenario, where most main sequence stars are converted to black holes late in the universe by dark matter-seeded collapse. We now discuss some other ways CNS might be realized. 

%%%%
\subsection{Alternative CNS measures and dynamics}
\label{subsec:variations}
%%%%

Thus far we have identified heavy particle dark matter with nuclear scattering interactions as the agent that brings about production of black holes, as sketched out in ~\eqref{eq:unific}.
Depending on the mass of dark matter, either the total mass or the number of black holes formed is maximized.
This scenario may be placed in the context of other frameworks that follow the broad principle of CNS, which is to form self-replicating universes that look like our own.
We collect these variations in Table~\ref{tab:othermeasures}, on which we elaborate now.

The simplest variation is demanding that, instead of the baryonic mass budget, most of the dark matter mass be used for making new universes.
Then one might simply predict that dark matter is made of primordial black holes (PBHs), or of star-like dark matter structures that collapse to form black holes, each containing a universe. 
If one wished to maximize the number of universes one might then require lighter PBHs.
On the other hand, one might wonder whether a universe may only form in a black hole if it weighs in the vicinity of a solar mass~\cite{MassUniverseBH:Poplawski:2011ip}. In that case, either (i) asymmetric particle dark matter is required to transmute stars to black holes, or (ii) PBHs in the unconstrained $10^{-16}-10^{-11}~M_\odot$ PBH mass window, which could be all the dark matter, may be captured in main sequence stars via dynamical friction and accrete stellar material to grow into heavy black holes~\cite{PBHsStars:TinyakovEsser:2022owk}. In the latter mechanism there would be a question (relative to heavy PBH formation) as to whether a universe can form in a black hole if a small black hole gains mass and crosses some universe-bearing threshold.

Returning to the assumption that the size of the black hole does not affect new universe formation, and assuming PBHs are typically formed at the end of inflation (and definitely before the recombination epoch to be consistent with cosmic microwave background measurements), it is not clear why a typical universe like ours must undergo all the subsequent epochs, notably galaxy and star formation.
On the other hand, our realization of CNS in \eqref{eq:unific} makes every epoch -- and particularly the stelliferous one -- significant, and unifies the timescale for universe formation with other dynamic timescales.

We can also consider what CNS has to say about the dark matter content of the universe.
The total dark matter mass is observed to be comparable to the baryonic mass, $\Omega_{\rm DM} \simeq 5.4\Omega_{\rm b}$.
In our treatment, only a small $\lesssim \Oc(10^{-8})$ fraction of the dark matter mass goes into converting stellar baryons to black holes in $10^{14}$~yr (as seen from Eq.~\eqref{eq:masscapturerate}), making it seem like the cosmic dark matter content is over-optimized for self-replication of universes. 
However, had $\Omega_{\rm DM} \ll \Omega_{\rm b}$, galaxy formation on small scales would have been suppressed without the gravitational influence of dark matter, and feedback effects from stellar winds and supernovae would have unbound large quantities of material from the earliest stellar clusters.
Further, the scarcity of heavy elements made in supernovae would hamper the cooling of gas clouds.
All these effects suppress conventional star formation and weight the IMF toward the high mass end, which would seem to work against making the most black holes within the apparent dynamics of our universe.
It is also unclear if $\Omega_{\rm DM} \gg \Omega_{\rm b}$ -- even if $\Omega_{\rm DM}$ is not high enough to overclose the universe -- is an optimal situation. 
This seemingly leads to more galaxy and star formation than in our universe, however baryonic feedback effects (stellar winds, supernovae, AGN feedback, etc.) in the first galaxies might actually suppress star formation over the long run.
The nature of baryonic feedback is highly uncertain and an ongoing field of research~\cite{SilkMamonReview,Vogelsberger:2019ynw,Acevedo:2023cab}.

Turning to a different universe production mechanism, if dark energy is the agent of cosmic self-replication, Vilenkin's far-future de Sitter universe nucleating black holes (see Appendix~\ref{app:CNS3predictions}) fulfills the requirement~\cite{CNSdS:Vilenkin:2006hq}. It is then unclear how a universe like ours, with interesting phenomena like atom and star formation, is typical. By contrast, in our CNS framework, we have shown that there could be a link between the major entropy-producing processes in our universe and BH/universe production.

This brings us to the question of whether non-stellar sources of entropy increase might be considered.
In our work we have used the entropy sourced by reprocessed starlight as the predominant non-horizon entropy source.
But far more entropy than this is sourced by horizons, $i.e.$ of both black holes and the cosmic horizon of the observable universe. 
As in Ref.~\cite{BoussoHarnikKribs:2007kq} we have not considered this horizon entropy in the main text, as its relevance to observers and the birth of universes is unclear.
If horizon entropy is considered for some reason, our universe as we know it is enough to convert most of the mass of entropy sources to new universes. 
This is because in $10^{33}$~yr most baryonic mass would have gone into making galactic mass black holes through dynamical accumulation onto the supermassive black holes (SMBHs) at the centers of galaxies~\cite{AdamsLaughlin:1996xe}. 
These eventual SMBHs would be the major sources of horizon entropy.
In that scenario, star formation plays a key role, but the role of dark matter is unclear.
But if it is part of the equation, models of dark matter seeding SMBHs~\cite{DMvSMBHs:Feng:2020kxv} become significant.
The black hole count may also be maximized in these scenarios.
In addition, one could have super-massive PBHs that grow faster than regular SMBHs and contributing to the majority of entropy, although these could only make up a small fraction of dark matter owing to constraints from their accretion~\cite{PBHsGreenKavanagh:2020jor}.
Also, these dark matter-as-black holes versions of CNS do not explain star formation.

%%%%
\subsection{Perspective on reccurrence}
%%%%%

To wrap up, we discuss measure problems and their relation to CNS. As a prior example of measure problems, let us consider Weinberg's work on understanding the cosmological constant. In formulating his anthropic bound on the CC, Weinberg had assumed that the fraction of baryons that ends up in galaxies is another parameter (in addition to the CC) that scans on some landscape of vacua, as the anthropically conditioned probability distribution is proportional to the number of observers. 
But he had raised the possibility that the number of baryons may itself scan~\cite{MultiverseWeinberg:2005fh},
in which case one cannot know that our universe is typical.
Analogously, one parameter that could scan in the CNS framework is the fraction of baryonic mass that converts to black holes. As we have noted at the outset of our discussion, scanning parameters in this way does not necessarily lead to increased universe fitness.

However, in the context of cosmologies whose future looks similar to their past, let us note that CNS appears to generically circumvent the Boltzmann brain problem. Over timescales of $\gg 10^{10^{66}}$~yr, thermodynamic fluctuations could create low-entropy states that mimic conscious observers with false memories of a briefer evolution of the universe~\cite{Carroll:2017gkl}.
The problem is that these ``Boltzmann brains" would far outnumber ``regular brains" formed through better known evolutionary pathways, implying we must ask whether we are the former.
The solution would be to construct a cosmology where Boltzmann brains are atypical.
In CNS as we have outlined, the formation of new universes occurs over $10^{14}$ year timescales, and given that the scaling of universe growth is exponential with time, the number of universes with regular-brain observers will be exponentially larger than Boltzmann brains.
This applies to both CNS1 and the revised version we have presented, the latter with an even larger profusion of black holes made in a relatively short $10^{14}$ yr period.
Note that eternal inflation, which is invoked as an alternative to CNS (Sec.~\ref{subsec:CNS}), generally leads to the Boltzmann brain problem~\cite{Albrecht:2004ke,Bousso:2006xc,Linde:2006nw,Carroll:2017gkl}.

%%%%
\section*{Acknowledgments}

We thank Lee Smolin for conversations and comments on this work, and Robert Brandenberger, Shantanu Desai and Aaron Vincent for discussions. This work was supported by the Arthur B. McDonald Canadian Astroparticle Physics Research Institute, the Natural Sciences and Engineering Research Council of Canada (NSERC), and the Canada Foundation for Innovation. Research at Perimeter Institute is supported by the Government of Canada through the Department of Innovation, Science, and Economic Development, and by the Province of Ontario.

\appendix

%%%%
\section{Critical commentary on cosmological natural selection}
\label{app:CNS3predictions}
%%%%%

In the original version of CNS, Smolin proposed three tests that were suggested as ways to falsify the hypothesis.
If these tests are taken at face value it could be argued that the hypothesis has been falsified.
In Section \ref{sec:discs} we have already discussed why, in our opinion, the predictions laid out in CNS1 are not necessarily ideal tests. Moreover, we think our findings in this work have provided some new ways to test the theory. Our major objection to the original tests is that the global fitness of the universe need not increase by naively varying a single parameter to increase black hole production, which is in agreement with results from evolutionary biology~\cite{Stearns2000}. 
Nevertheless, we think it is instructive to review the predictions laid out in Ref.~\cite{Smolin:1990us}, and comment on their status vis-a-vis current astrophysical and cosmological data. The CNS1 predictions advocated in Ref.~\cite{Smolin:1994vb,Smolin:1997pe} are as follows:

(i) Neutron stars must be composed of kaon-condensates, which soften the high-density equation of state and hence lower the maximum mass of stable NSs, $M^{\rm NS}_{\rm max}$.
The smaller $M^{\rm NS}_{\rm max}$ is, the wider the range of progenitor masses that end up as black holes as opposed to NSs. 
The strange quark mass is then optimal in our universe\footnote{Apart from this prediction, CNS may also explain the existence of the heavier fermion generations as relics of early (literal) generations of self-reproducing universes mutating toward a set of fermion masses optimal for production of stars and BHs~\cite{NambuprotoCNS:1985tt,Smolin:1990us}.} as variations in it could vary $M^{\rm NS}_{\rm max}$ without influencing massive star formation and supernovae. 
In Ref.~\cite{Smolin:2006gt} $M^{\rm NS}_{\rm max}$ is predicted as about 1.6 $M_\odot$.  

(ii) Cosmic inflation must be determined by a single inflaton coupling parameter which controls both density fluctuations $\delta \rho/\rho$ and  the number of $e$-foldings $N_e$.
This way, increasing $\delta \rho/\rho$ -- which has the effect of increasing PBH formation -- would also result in decreasing $N_e$ -- which exponentially reduces the volume of the universe and thus PBH population. 
If inflation were discovered to be described by parameters that independently control $\delta\rho/\rho$ and $N_e$, CNS would be falsified. 

(iii) Early star formation must be suppressed.
Organic chemistry is required for the cooling of star-forming molecular hydrogen gas clouds from which black hole progenitors arise.
In particular, radiation from molecular vibrations of CO dominantly cool the gas clouds, and carbon dust and ice shield the clouds from ultraviolet light emitted by massive stars.
If an alternative channel existed, it would would have operated at high redshifts when carbon and oxygen abundances were much smaller, and produced many more core-collapse supernovae than observed.
Thus the neutron-proton mass difference, being smaller than nuclear binding energies in our universe, is selected to pave the way for star formation but only at $t > 10^6$~yr.

Now we provide our remarks.
Prediction (i), on the face of it, is in tension with the observation of NSs heavier than 2~$M_\odot$; the heaviest at the time of writing is estimated to weigh 2.35~$\pm$~0.17~$M_\odot$~\cite{NSFastestHeaviest:Romani:2022jhd}.
This prediction/bound for heavy NSs was in any case blurry.
The extra span in mass range of black holes made by core-collapse of main-sequence stars to kaon-condensate (or hyperon) stars would arguably result in a small gain in black holes formed, compared to the total range of astrophysical black hole masses, $\Oc(10^0-10^2)~M_\odot$, which is further impacted by the effects of stellar metallicity~\cite{BHmasspectrum-Spera-2015,BHmassfunc:Sicilia:2021gtu}. Moreover, it is not clear that nature's actual dynamics for forming black holes and NSs in core collapse supernovae would result in more black holes being formed if the maximum NS mass is lowered. There is a deficit (``mass gap") observed in 2$-$5 solar mass compact objects in x-ray binary systems~\cite{Farr_2011}, which is in tension with models of core collapse supernovae that predict a continuous population of NSs/black holes produced in this mass range. If core collapse supernovae, owing to the still-emerging details of their dynamical collapse, do not form compact objects in this mass range, then the maximum mass of NSs -- itself set by the unknown high-density equation of state of nuclear matter -- may be irrelevant for determining the number of black holes produced.

Prediction (ii) should be contrasted with currently favored models of inflation that permit inflationary potentials with technically a single coupling parameter for the inflaton ($e.g.$, hilltop inflation)~\cite{Planck:2018jri}.
In these models larger density fluctuations are associated with {\em more} $e$-folds of inflation, since these are so-called ``Planck flat" models~\cite{Bramante:2014rva}. Regardless, we think the existence of additional dynamical pathways to form black holes is not a problem for CNS as explained above. We rather do expect multiple black hole-forming pathways, and to us what is most interesting to consider is whether the dominant form of entropy production in the universe can be linked to black hole production.

Prediction (iii) is on uncertain footing. 
As is well known and pointed out in a critique of CNS1 in Ref.~\cite{SilkHolisticCosmology}, at no epoch is organic chemistry needed for star formation, which may be simply triggered by cooling and collapse of massive hydrogen gas clouds.
Further, copious star formation at very early times may have already been observed at the James Webb Space Telescope, which detected bright and active galaxies at and above unexpectedly high redshifts of 10 or so (see the references in Ref.~\cite{sun2023bursty}).
This has motivated studies on potential pathways to early star formation, none of which particularly rely on carbon-oxygen chemistry~\cite{Robertson:2022gdk,Dekel2023,Qin:2023rtf,trinca2024exploring,sun2023bursty,Klessen:2023qmc}.

Finally, we discuss one notable objection to the paradigm of CNS, which was raised by Vilenkin~\cite{CNSdS:Vilenkin:2006hq}.
In the far future, when the universe may be described by de Sitter space, semiclassical descriptions of gravity with instantons imply that black holes will be nucleated via quantum tunneling. 
The nucleation rate goes as $\exp[-M_{\rm BH}/T]$, with $T = H/2\pi = \sqrt{\Lambda_{\rm cc}/3}/2\pi$ the Gibbons-Hawking temperature of the quantum fields in de Sitter space, where $H$ is the Hubble rate and $\Lambda_{\rm cc}$ is the CC.
Thus by increasing $\Lambda_{\rm cc}$ to more than the observed value, the number of black holes formed is exponentially increased, apparently falsifying CNS.
Smolin's response~\cite{Smolin:2006gt} includes the arguments that currently known physics cannot be extrapolated to much vaster scales in distance, time, and energy, 
that CNS concerns itself with explaining the universe as we observe it today,
and that random fluctuations can always produce entities that outnumber their equivalents formed through an ordering principle, e.g., DNA sequences formed outside reproducing organisms and Poincare recurrence-induced Boltzmann brains, due to which care must be taken in choosing the ensemble in which we want to show that an entity is ``typical".
(These arguments also apply to the criticism of CNS that blue-tilting the spectral index of primordial fluctuations would enhance black hole production~\cite{SilkHolisticCosmology}.)

We remark that our version of CNS provides a different response to Vilenkin's late universe fluctuation production.
Whereas in Ref.~\cite{CNSdS:Vilenkin:2006hq} one had to wait colossal timescales (of about $10^{71}$~yr) for universe reproduction to transpire, in dark matter-induced CNS this is achieved over the relatively short time of $10^{14}$~yr.
Hence although a second route exists for making copious black holes, it may be that our universe follows a different route for optimal reproduction (see also Section \ref{sec:discs}).
This has the added advantage of not having to extrapolate known physics to exceptionally long timescales, as also argued in Ref.~\cite{Smolin:2006gt}.

\bibliography{main}

\end{document}